\documentstyle[11pt,paspconf,epsf]{article}

\begin{document}

\title{Results on Galaxy Evolution from the 
       CNOC2 Field Galaxy Redshift Survey}

\author{H. Lin\altaffilmark{1}, 
        H.K.C. Yee\altaffilmark{1}, 
        R.G. Carlberg\altaffilmark{1}, 
        S.L. Morris\altaffilmark{2},
        M. Sawicki\altaffilmark{1}, 
        D.R. Patton\altaffilmark{3}, 
        G.D. Wirth\altaffilmark{3}, 
        C.W. Shepherd\altaffilmark{1},
        E. Ellingson\altaffilmark{4},
        D. Schade\altaffilmark{2}, 
        R.O. Marzke\altaffilmark{5}, \&
        C.J. Pritchet\altaffilmark{3}}

\altaffiltext{1}{Department of Astronomy, University of Toronto,
   Toronto, ON M5S 3H8, Canada}
\altaffiltext{2}{Dominion Astrophysical Observatory, Herzberg Institute of
   Astrophysics, Victoria, BC V8X 4M6, Canada}
\altaffiltext{3}{Department of Physics \& Astronomy, University of Victoria,
   Victoria, BC V8W 3P6, Canada}
\altaffiltext{4}{Center for Astrophysics \& Space Astronomy, University of
   Colorado, Boulder, CO 80309, USA}
\altaffiltext{5}{Carnegie Observatories, Pasadena, CA 91101, USA}

\begin{abstract}
The CNOC2 Field Galaxy Redshift Survey presently contains some 5000
galaxy redshifts, plus extensive $U \! BgRI$ photometry, and is the
largest galaxy sample at moderate redshifts $0.1 < z < 0.6$. Here we
present some {\it preliminary} results on the galaxy luminosity
function (LF) and its redshift evolution, using a 
sample of $R < 21.5$ CNOC2 galaxies, subdivided into
early, intermediate, and late types based on their $B-R$ colors
relative to {\it non-evolving} galaxy models. We find a significant
steepening in the faint-end slope $\alpha$ of the LF as one proceeds
from early to late types. Also, for all galaxy types 
we find a rate of $M^*$ evolution consistent with that from 
passively evolving galaxy models. Finally, late-type galaxies show positive
density evolution with redshift, in contrast to negative or no density
evolution for earlier types. 
\end{abstract}

%\keywords{}

\section{Introduction to the Survey}

The determination of the clustering, luminosity, star formation and
other properties of galaxies, and the measurement of the evolution
of those properties with redshift, are fundamentally important goals
of observational cosmology. The CNOC2 (Canadian Network for
Observational Cosmology) Field Galaxy Redshift Survey is explicitly
designed to study the evolution of galaxy clustering and luminosity
properties over $0.1 < z < 0.6$, and currently
constitutes the largest galaxy sample at these intermediate redshifts 
(see Yee et al.\ 1997 for more details). 
The survey has a total area of some 1.5 deg$^2$, 
in the form of 4 widely-separated patches on the sky, where each patch
is an ``L''-shape that spans about 1.5$^\circ$. The spectroscopic observations 
are carried out at the Canada-France-Hawaii Telescope, using efficient MOS
multi-slit spectroscopy of typically 100 objects at once. This is made 
possible with a band-limiting filter that restricts the spectra to 
4300-6300 \AA, resulting in a redshift range $0.12 < z < 0.55$ with
unbiased coverage of important galaxy absorption and emission
features. Redshift errors are about 75 km s$^{-1}$ in the rest frame. 
The survey also includes 5-color $U \! BgRI$ photometry for
nearly all fields (plus $K$ for some), with average 5$\sigma$ limits
at (Kron-Cousins) $R = 24.0$ and $B = 24.6$. 

\begin{figure}
\plotfiddle{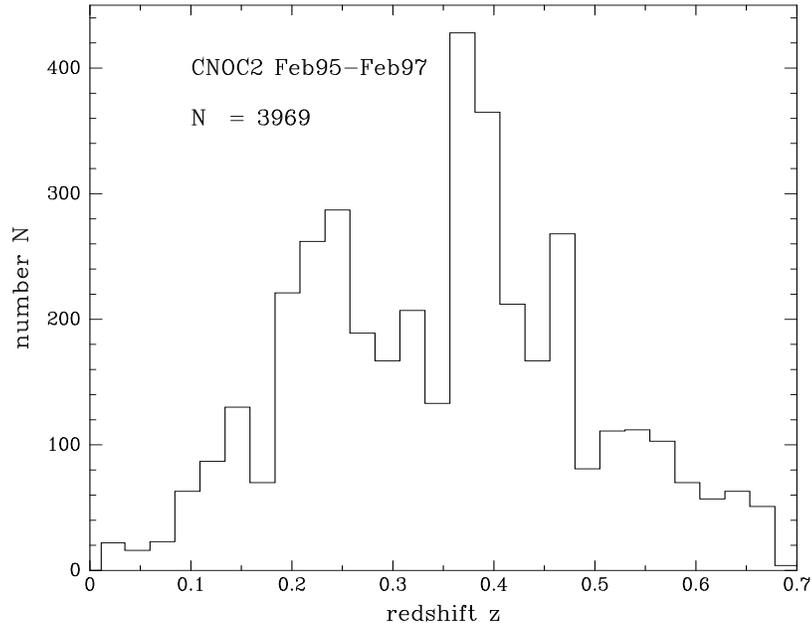}{3in}{90}{45}{45}{175}{-20}
\caption{The present CNOC2 survey redshift histogram. Note that another 1000
  redshifts have been obtained but are not yet reduced.} \label{fighist}
\end{figure}

\begin{figure}[t]
\plotfiddle{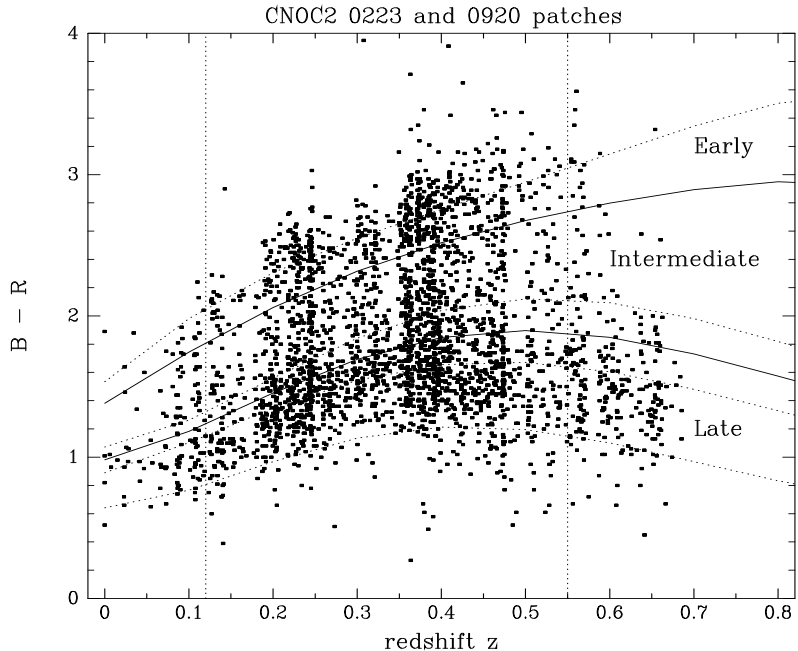}{3in}{90}{45}{45}{175}{-20}
\caption{$B-R$ color vs.\ $z$ for CNOC2 galaxies.
  Dotted lines show $k$-corrected $B-R$ from Coleman, Wu, \&
  Weedman (1980) SED's, and solid lines show the interpolations used to
  subdivide CNOC2 galaxies. The vertical lines denote the redshift range $0.12
  < z < 0.55$, within which there is unbiased coverage of
  spectral features important to redshift identification 
  for all galaxy types.
  } \label{figcol}
\end{figure}

At present (11/97) the CNOC2 survey is 80\% complete, with 3 of the 4
patches essentially done. The survey contains
some 5000 redshifts, with nearly 4000 for galaxies brighter than the 
nominal $R = 21.5$ spectroscopic completeness limit, at which the
cumulative redshift sampling rate is about 1 in 2.
Figure~\ref{fighist} shows the redshift
histogram for the survey, and Figure~\ref{figcol} plots observed $B-R$
color vs.\ redshift. In the present paper we
discuss some {\it preliminary} results on the evolution of the galaxy
luminosity function, using a subset of about 2000 $R < 21.5$ galaxies 
from the ``0223'' and ``0920'' patches, which are the 2 current
patches with almost fully reduced redshift and 5-color data.
We adopt $H_0 = 100 \ h$ km s$^{-1}$ Mpc$^{-1}$ and $q_0 = 0.5$
throughout, as is typically done in the analysis of other redshift 
surveys at similar $z$.

\section{The Luminosity Function}

The luminosity function (LF) of galaxies is a fundamentally
important quantity in the study of galaxy populations and their
evolution. Accurate determinations of
the LF at both low and high redshifts are crucial.
Large wide-angle redshift surveys are providing precise measurements
of the luminosity function in the local $z \sim 0$ universe (e.g.
Loveday et al.\ 1992; Marzke et al.\ 1994b; Lin et al.\ 1996),
and smaller but deeper surveys have begun to measure
the LF up to redshifts $z \sim 1$
(e.g., Lilly et al.\ 1995; Ellis et al.\ 1996; Cowie et al.\ 1996;
Lin et al.\ 1997), revealing
clearly the rapid evolution of the LF of blue, star-forming galaxies, but
relatively little change in the LF of red, more quiescent objects.
Here we will examine the luminosity function and its evolution for 
galaxies of different types in the CNOC2 sample, and we will exploit 
the large CNOC2 sample size to derive improved constraints.

\begin{figure}
\plotfiddle{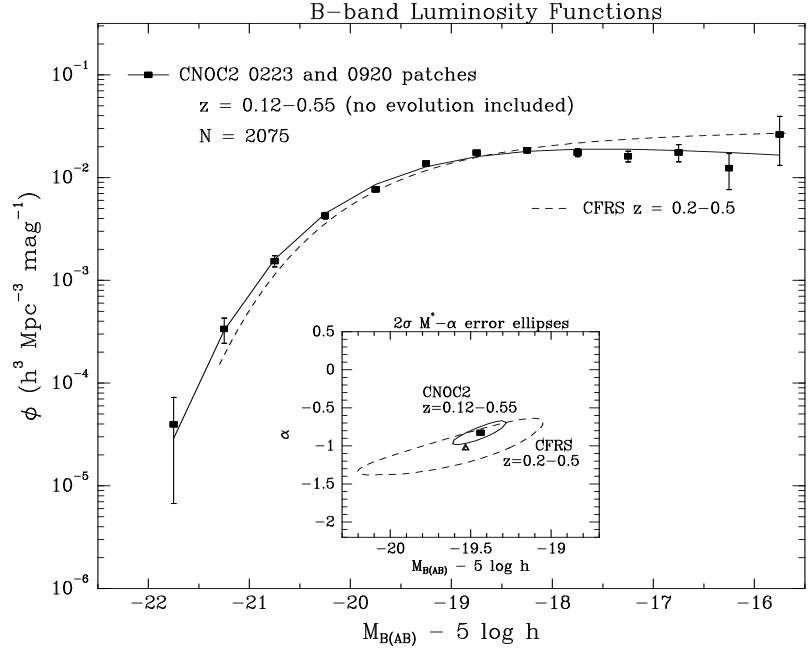}{3in}{90}{45}{45}{175}{-20}
\caption{Comparison of the CNOC2 $B_{AB}$-band luminosity function
  with that from the much smaller CFRS sample.} \label{figlfs}
\end{figure}

\begin{figure}
\plotfiddle{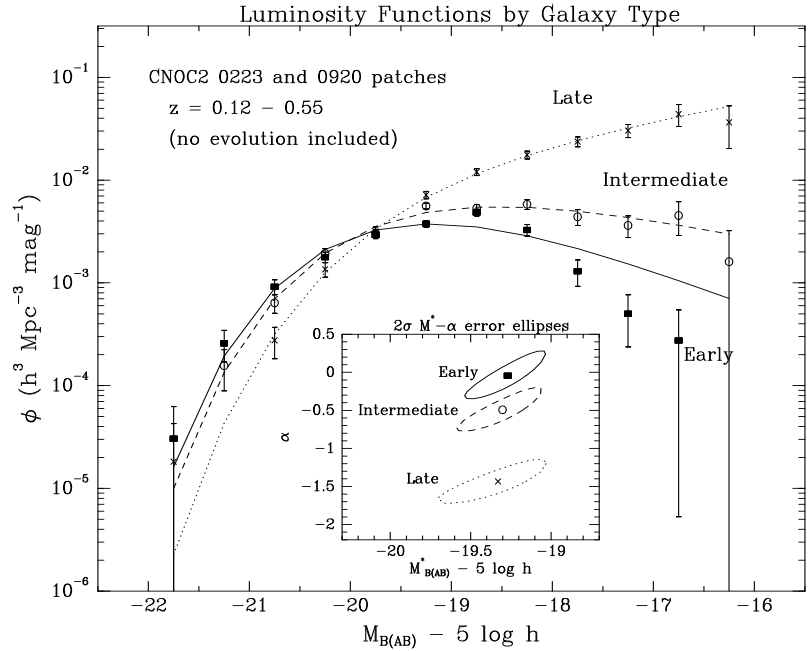}{3in}{90}{45}{45}{175}{-20}
\caption{Comparison of $B_{AB}$-band luminosity functions for CNOC2
  galaxies of different types, classified by observed $B-R$ color as
  shown in Figure~\ref{figcol}.} \label{figlfrb}
\end{figure}

Figure~\ref{figlfs} shows the $B_{AB}$-band luminosity function for
a sample of 2075 galaxies, with $R < 21.5$ and $0.12 < z < 0.55$, 
from the CNOC2 0223 and 0920
patches. The LF is computed using standard inhomogeneity-independent
maximum-likelihood
techniques (e.g., Efstathiou, Ellis, \& Peterson 1988). The CNOC2 LF
appears to be well fit by a nearly-flat Schechter function 
($\alpha = -0.8$), and it is also consistent with the $0.2 < z < 0.5$ LF 
from the Canada-France Redshift Survey (CFRS; Lilly et al.\ 1995), as seen
in the comparison of 2$\sigma$ $M^*$-$\alpha$ error ellipses in the
Figure~\ref{figlfs} inset. Note in particular the much smaller LF uncertainties
of the CNOC2 sample, which contains about ten times as many galaxies
as the CFRS over the same redshift range.

However, as has been observed in many previous galaxy samples, the
luminosity function does appear to depend on various galaxy properties,
such as rest-frame color, spectral type, or morphology (see e.g., 
Lilly et al.\ 1995; Lin et al.\ 1996, 1997; Heyl et al.\ 1997; 
Marzke et al.\ 1994a). 
To check the type dependence of the LF, we have also divided CNOC2
galaxies into early, intermediate, and late types (more precisely red,
intermediate, and blue), based on
$k$-corrected $B-R$ colors computed from non-evolving Coleman, Wu, \&
Weedman (1980; CWW) spectral energy distributions (SED's; Figure~\ref{figcol}).
Figure~\ref{figlfrb} then clearly shows that the luminosity function depends
on galaxy type, with a strong steepening of the faint-end slope
$\alpha$ as one proceeds from red, early-type galaxies to blue,
late-type ones. These type-dependent differences are also quite
significant, as shown in the Figure~\ref{figlfrb} inset, and are
also similar to those seen in previous 
smaller surveys at moderate redshifts, including the CFRS (Lilly et
al.\ 1995) and Autofib (Ellis et al.\ 1996) samples.

\section{Luminosity and Number Density Evolution}

\begin{figure}
\plotfiddle{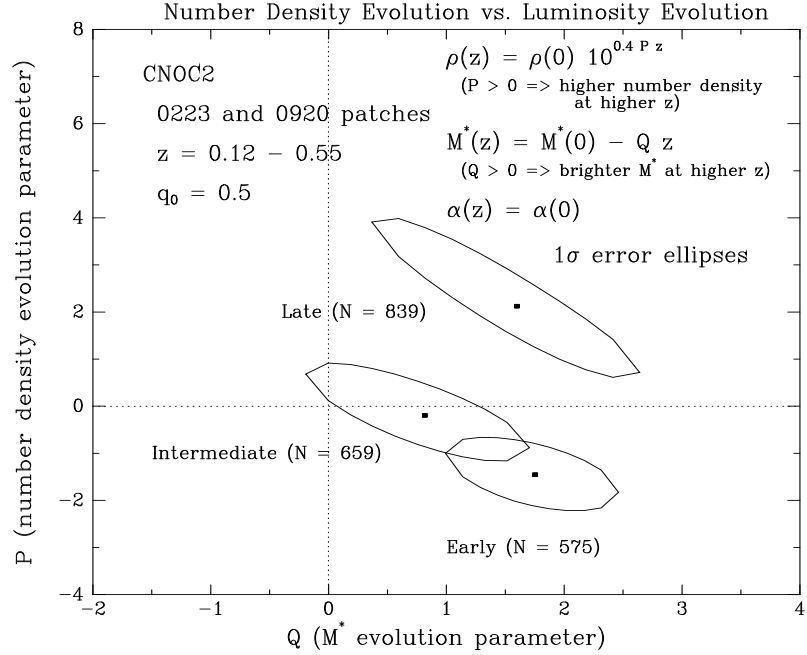}{3in}{90}{45}{45}{175}{-20}
\caption{1$\sigma$ error ellipses in $P$ (the number density evolution
  parameter) vs.\ $Q$ (the $M^*$ evolution parameter), for
  different CNOC2 galaxy types.} \label{figdep}
\end{figure}

So far the LF's have been fit without regard to evolution, but we know
from previous surveys that galaxies do evolve significantly over $z <
1$. As a simple but useful parameterization of luminosity and number
density evolution, we adopt the following model:
\begin{eqnarray}
\rho(z) & = & \rho(0) 10^{0.4 P z} \nonumber \\ 
M^*(z) & = & M^*(0) - Q z \\
\alpha(z) & = & \alpha(0) \nonumber
\end{eqnarray} 
where $\rho$ is the galaxy number density, $P$ quantifies the rate of
number density evolution, and $Q$ describes the rate of $M^*$ or
luminosity evolution. $P$ and $Q$ are essentially just the linear
coefficents in an expansion in $z$ of $\rho$ and $M^*$, respectively.
For simplicity, $\alpha$ is assumed not to evolve so that the LF shape
does not change with redshift. We use maximum-likelihood techniques 
to derive best-fitting LF and evolution parameters, and 
Figure~\ref{figdep} shows the resulting 1$\sigma$ $P$ vs.\ $Q$ error
ellipses for the three CNOC2 galaxy types, which do indeed show
different evolutionary trends. In particular, late-type galaxies show
positive density evolution ($P > 0$ so $\rho$ increases at higher $z$),
early-type galaxies show negative density evolution ($P < 0$), and
intermediate types show no density evolution. On the other hand, all
three galaxy types show positive luminosity evolution, i.e.\ $M^*$ is
brighter in the past, and the derived values of $Q \approx 1-2$ are
consistent with those expected from passively evolving galaxy models
(e.g., Bruzual \& Charlot 1993). (Note that using $q_0 = 0.1$ instead
of $q_0 = 0.5$ will roughly decrease $P$ by 0.8 and increase $Q$ by 0.4.)

\begin{figure}
\plotfiddle{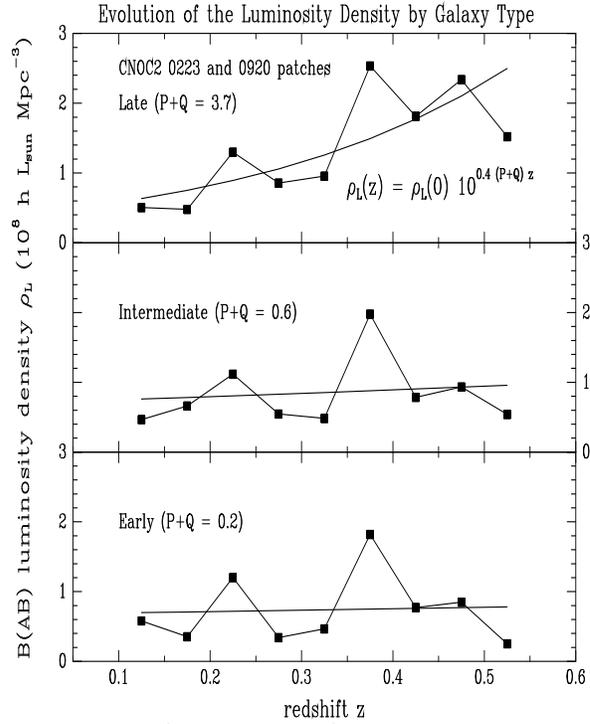}{3in}{0}{40}{40}{-125}{-40}
\caption{The redshift evolution of the $B_{AB}$-band luminosity
  density $\rho_L$ for different CNOC2 galaxy types.} \label{figld}
\end{figure}

\begin{figure}[t]
\plotfiddle{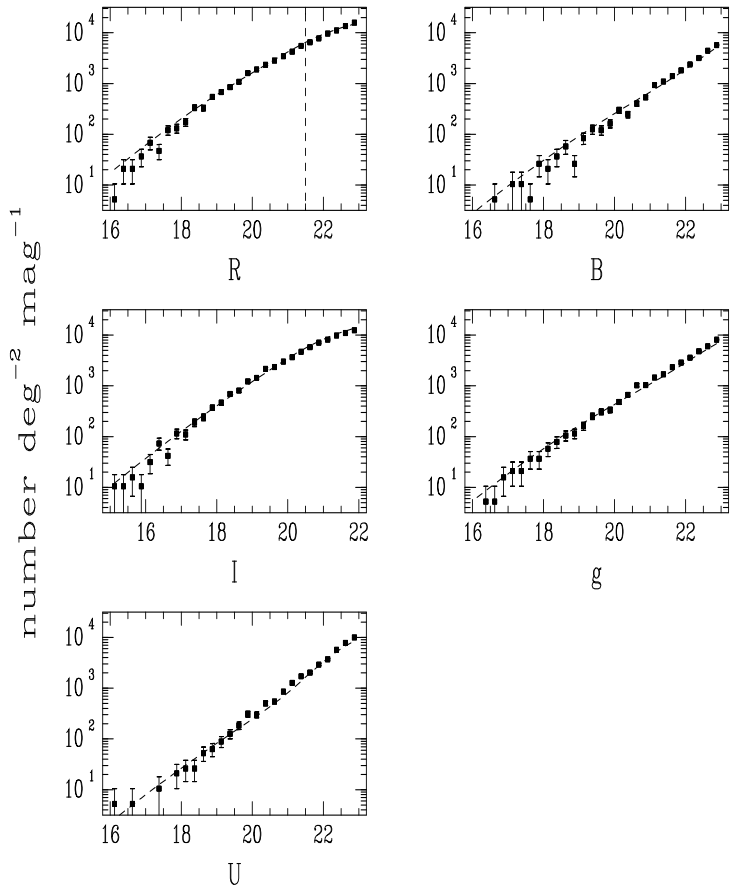}{4in}{0}{50}{50}{-150}{-40}
\caption{Comparison of observed CNOC2 galaxy counts (points),
  in the $R$, $B$, $I$, $g$, and $U$ bands, with those computed from
  the LF evolution model fit to the spectroscopic sample (lines). Note that the
  spectroscopic sample used to derive the LF evolution parameters 
  is limited to $R < 21.5$, as indicated by the vertical line in the
  $R$ panel.} \label{figmag}
\end{figure}

Also, in our parameterization,
the luminosity density varies as $\rho_L(z) = \rho_L(0) 10^{0.4 (P+Q)
z}$, so $P+Q$ characterizes the rate of evolution of the luminosity
density. Figure~\ref{figld} plots $\rho_L$ vs.\ $z$ for the three
galaxy types and shows the rapid rise in the luminosity density of
late-type galaxies relative to that of early- and intermediate-type
galaxies, again consistent with similar trends seen in previous
smaller surveys (e.g., Lilly et al.\ 1996; Lin et al.\ 1997). Finally,
Figure~\ref{figmag} compares the observed CNOC2 galaxy number counts 
with those predicted by the LF and evolution parameters fit from the
spectroscopic data {\it alone}. The observed and LF-predicted counts are in good
agreement for all 5 CNOC2 bands, showing that our simple 3-population 
evolution model is a
reasonable description, and that there are no conspicuous unaccounted
spectroscopic selection effects. Note also from the $R$ counts that 
our LF evolution model remains valid beyond the adopted $R = 21.5$
limit of the spectroscopic sample. Moreover, as our galaxy
classifications are based only on observed $B-R$ colors, 
the good match to the $I$, $U$, and $g$ counts gives us an independent check
on the validity of our LF models.

However, an important caveat needs to be kept in mind, namely the
sensitivity of the LF evolution results to the precise choice of 
SED's used to classify galaxies. In particular, the present choice of 
non-evolving SED's obviously does not account for the color evolution
of galaxies. If one were to use more physically motivated {\it evolving} 
SED's, the resulting galaxy classifications will generally be 
different. For example, one can eliminate the negative density
evolution of early-type galaxies by using simple evolving models with
exponentially declining star formation rates (Bruzual \& Charlot 1993), 
wherein galaxies are bluer at higher redshifts; this allows more
galaxies to be classified as early-type at higher $z$, and
thus increases the value of $P$. There are obviously many
possible combinations of evolving SED's that one may choose, by varying
parameters such as star formation rate, age, metallicity, or dust content.
Work is ongoing to test particular evolutionary scenarios, for example, 
to see whether
simple pure luminosity evolution models (no density evolution), with
various choices of star formation history, age, etc.,
are consistent with both the CNOC2 LF and number count data. (The
availability of 5-color data for all CNOC2 galaxies will be of great
help here, as it permits tight constraints on the SED of any
particular galaxy.) The LF evolution model outlined in
the present paper should thus be considered as a first step, a 
{\it description} of galaxy evolution within the
framework of non-evolving SED's, rather than 
an {\it explanation} of galaxy evolution in terms of more physically
motivated evolving models.

We will also be able to constrain those physical models using
additional tools. Note that there is much 
spectroscopic information available, which will permit measurements of
star formation rates using emission line indices, as well as 
classification of galaxies via principal component analysis. Moreover,
photometric redshifts can be accurately calibrated using the 
spectroscopic sample, and use of a larger photometric-redshift sample
will help to constrain the LF at fainter magnitudes, as well as
reduce errors on the $P$ and $Q$ parameters. Measurements of the evolution of the
surface brightnesses and morphologies of CNOC2 galaxies will
also be possible. Finally, the environmental dependence of luminosity and star
formation properties, and the relation to the evolution of galaxy
clustering, will be explored using catalogs of groups
and pairs of galaxies, and through measurements of the 
correlation function by galaxy type. Ultimately we hope to put
clustering, luminosity, star formation, and spectral
properties together into a coherent and self-consistent 
picture of galaxy evolution at moderate redshifts $z < 1$.

%\acknowledgments

\end{document}